\documentclass[english,aps,prl,twocolumn]{revtex4-1}
\usepackage[T1]{fontenc}
\usepackage[latin9]{inputenc}
\usepackage{color}
\usepackage{textcomp}
\usepackage{amsmath}
\usepackage{amssymb}
\usepackage{graphicx}

\makeatletter
 \@ifundefined{textcolor}{}
 {%
  \definecolor{BLACK}{gray}{0}
  \definecolor{WHITE}{gray}{1}
  \definecolor{RED}{rgb}{1,0,0}
  \definecolor{GREEN}{rgb}{0,1,0}
  \definecolor{BLUE}{rgb}{0,0,1}
  \definecolor{CYAN}{cmyk}{1,0,0,0}
  \definecolor{MAGENTA}{cmyk}{0,1,0,0}
  \definecolor{YELLOW}{cmyk}{0,0,1,0}
  }

\makeatother

\usepackage{babel}
\begin{document}

\title{Dynamics of broken symmetry nodal and anti-nodal excitations in Bi$_{2}$Sr$_{2}$CaCu$_{2}$O$_{8+\delta}$
probed by polarized femtosecond spectroscopy}

\author{Y. Toda}

\affiliation{Department of Applied Physics, Hokkaido University, Sapporo 060-8628,
Japan.}

\author{F. Kawanokami}

\affiliation{Department of Applied Physics, Hokkaido University, Sapporo 060-8628,
Japan.}

\author{T. Kurosawa}

\affiliation{Department of Physics, Hokkaido University, Sapporo 060-0810, Japan.}

\author{M. Oda}

\affiliation{Department of Physics, Hokkaido University, Sapporo 060-0810, Japan.}

\author{I. Madan}

\affiliation{Complex Matter Dept., Jozef Stefan Institute, Jamova 39, Ljubljana,
SI-1000, Slovenia}

\author{T. Mertelj}

\affiliation{Complex Matter Dept., Jozef Stefan Institute, Jamova 39, Ljubljana,
SI-1000, Slovenia}

\author{V. V. Kabanov}

\affiliation{Complex Matter Dept., Jozef Stefan Institute, Jamova 39, Ljubljana,
SI-1000, Slovenia}

\author{D. Mihailovic}

\affiliation{Complex Matter Dept., Jozef Stefan Institute, Jamova 39, Ljubljana,
SI-1000, Slovenia}

\date{\today}
\begin{abstract}
The dynamics of excitations with different symmetry is investigated
in the superconducting (SC) and normal state of the high-temperature
superconductor Bi$_{2}$Sr$_{2}$CaCu$_{2}$O$_{8+\delta}$ (Bi2212)
using optical pump-probe (Pp) experiments with different light polarizations
at different doping levels. The observation of distinct selection
rules for SC excitations, present in A$_{{\rm 1g}}$ and B$_{{\rm 1g}}$ symmetries,
and for the PG excitations, present in A$_{{\rm 1g}}$ and B$_{{\rm 2g}}$ symmetries,
by the probe and absence of any dependence on the pump beam polarization
leads to the unequivocal conclusion of the existence of a spontaneous
spatial symmetry breaking in the pseudogap (PG) state. 
\end{abstract}


\maketitle
Ultrafast pump-probe (Pp) spectroscopy has been widely used to investigate
the high-$T_{\mathrm{c}}$ superconductivity from various viewpoints
\cite{Kaindl05,Perfetti07,Fausti11,Graf11,Giannetti11}. Nonequilibrium
studies give a unique insight into quasiparticle (QP) dynamics, revealing
universal two-component QP dynamics associated with the superconducting
(SC) gap and pseudogap (PG) excitations in high-$T_{\mathrm{c}}$
materials. The two types of excitations were characterized by distinct
relaxation times, temperature dependencies, and/or sign of the optical
signal, depending on the material, doping level, photo-excitation
intensity and the wavelengths of light used in the Pp experiments \cite{Demsar99,Kusar05,Liu08,Stojchevska11,Coslovich13}.
The dependence on the probe photon polarization of the two-component
reflectivity dynamics has also been reported \cite{Dvorsek02}. However,
the absence of a fundamental understanding of the optical processes
involved in Pp experiments so far prevented analysis of the symmetry
of excitations or detailed theoretical analysis of the excitations
on a microscopic level. Here, by performing a concise symmetry analysis
of Pp experiments on Bi2212 high temperature superconductors and identifying
the processes involved we open the way to investigations of the dynamics
of states associated with hidden broken symmetry and local or mesoscopic
symmetry breaking in systems with competing orders.

Generally, Pp experiments can be described as a two step process.
In the first step, the $pump$ pulse excitation can be viewed as a
process, which can be divided into a coherent stimulated-Raman (SRE)
excitation \cite{Stevens02} and an incoherent dissipative excitation
(DE). In the DE process, the high energy (eV) photoexcited carriers
create incoherent excitations by inelastic scattering on the timescale
of tens of fs resulting in a transient nonequilibrium occupation of
phonons (magnons, etc.) as well as a transient nonequilibrium QP occupation
$\Delta f$ near the Fermi level. The information about the incoming
photon polarization is lost during this process. In the SRE process,
various degrees of freedom are excited with a force $F(t)\propto P(t)$,
where $P(t)$ is related to the temporal profile of the pump pulse.
The symmetry of the coupling to different degrees of freedom is described
via the appropriate Raman tensor $\pi^{R}$ \cite{Stevens02}. For
the nonsymmetric modes the process should show characteristic dependence
on the pump photon polarization.

In pseudo-tetragonal ($D_{4h}$) symmetry, considered appropriate
for the cuprates \footnote{Conventionally the $x$ an $y$ axes are chosen along the Cu-O bond
directions.},
A$_{{\rm 1g}}$ and A$_{{\rm 2g}}$ as well as B$_{{\rm 1g}}$ and B$_{{\rm 2g}}$ excitations
can be coherently excited by the SRE process for the photon polarizations
lying in the CuO$_{2}$ plane \cite{Devereaux07}. Totally symmetric
A$_{1g}$ excitation such as DE can not coherently excite nonsymmetric
modes. However, \emph{an additional possibility exists}, where in the presence
of a local, dynamic or hidden symmetry breaking nonsymmetric modes
can be excited coherently also by the totally symmetric DE. This allows
us to probe symmetry breaking by means of the Pp spectroscopy.
\begin{figure}[h]
\includegraphics[width=0.98\columnwidth]{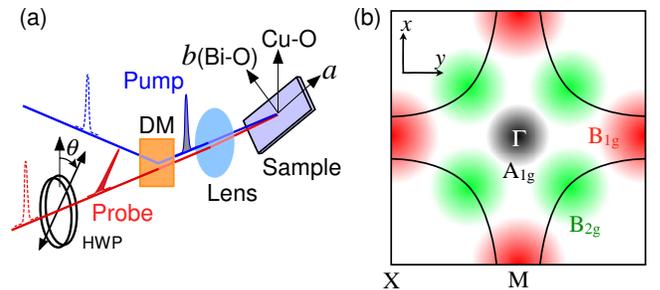} \caption{(a) A schematic illustration of the two-color pump-probe setup for
polarization-resolved measurements. The probe ($\lambda$=800 nm)
was variably polarized by a half-wave plate (HWP) and was combined
with the pump ($\lambda$=400 nm) by a dichroic mirror (DM). The angle
$\theta$ is measured relative to the Cu-O bond axes. (b) The $k$-space
selectivity of the probe according to the Raman-like process is indicated
(from \cite{Devereaux07}). }

\label{fig_temp} 
\end{figure}

In the second step of the Pp experiment, the transient change of reflectivity
detected by the probe can be described by the Raman-like process \cite{Stevens02}.
For QP excitations the polarization $P_{k}$ due to a pump-induced
change of $\Delta f(\mathbf{q})$ is given by: $P_{k}^{\mathbf{q}}=\sum_{l}\mathcal{R}_{kl}^{\mathbf{q}}E_{l}\Delta f(\mathbf{q})$,
where $\mathcal{R^{\mathbf{q}}}_{kl}=\frac{\partial\epsilon_{kl}}{\partial f(\mathbf{q)}}$
is a Raman-like tensor, $\epsilon_{kl}$ is the complex dielectric
tensor, and $E_{l}$ is the $l$-th component of the probe-pulse electric
field.

Assuming the pseudo-tetragonal structure ($D_{4h}$ point group) for
Bi2212, the photoinduced changes of the in-plane dielectric tensor
components can be decomposed according to symmetry as \footnote{A$_{2g}$ symmetry is omitted because it does not contribute in our experimental configuration.
}:
\begin{widetext} 
\begin{equation}
\Delta\mathbf{\epsilon}=\left[\begin{array}{cc}
\Delta\epsilon^{\mathrm{A}_{1\mathrm{g}}}\\
 & \Delta\epsilon^{\mathrm{A}_{1\mathrm{g}}}
\end{array}\right]+\left[\begin{array}{cc}
\Delta\epsilon^{\mathrm{B}_{1\mathrm{g}}}\\
 & -\Delta\epsilon^{\mathrm{B}_{1\mathrm{g}}}
\end{array}\right]+\left[\begin{array}{cc}
 & \Delta\epsilon^{\mathrm{B}_{2\mathrm{g}}}\\
\Delta\epsilon^{\mathrm{B}_{2\mathrm{g}}}
\end{array}\right].
\end{equation}
\end{widetext}Taking the in-plane probe incident electric field as
$E=E_{0}\bigl(\begin{smallmatrix}\cos\theta\\
\sin\theta
\end{smallmatrix}\bigr)$, then to the lowest order the angle-dependence of the photoinduced
change of reflectivity $R$ reduces to:

\begin{eqnarray}
\Delta R(\theta) & = & \frac{\partial R}{\partial\epsilon_{1}}\left[\Delta\epsilon_{1}^{\mathrm{A}_{1\mathrm{g}}}+\Delta\epsilon_{1}^{\mathrm{B}_{1\mathrm{g}}}\cos(2\theta)+\Delta\epsilon{}_{1}^{\mathrm{B}_{2\mathrm{g}}}\sin(2\theta)\right]+\nonumber \\
 &  & \frac{\partial R}{\partial\epsilon_{2}}\left[\Delta\epsilon_{2}^{\mathrm{A}_{1\mathrm{g}}}+\Delta\epsilon_{2}^{\mathrm{B}_{1\mathrm{g}}}\cos(2\theta)+\Delta\epsilon{}_{2}^{\mathrm{B}_{2\mathrm{g}}}\sin(2\theta)\right].\label{eq:poldep}
\end{eqnarray}
Here $\theta$ is defined in Fig. 1(a), and $\epsilon_{1}$ and $\epsilon_{2}$
are the real and imaginary parts of the in-plane dielectric constant,
respectively. Rewriting (\ref{eq:poldep}) in a more compact form
the angle-dependence of the transient reflectivity is:

\begin{equation}
\Delta R(\theta)\propto\Delta R_{\mathrm{A}_{1\mathrm{g}}}+\Delta R_{\mathrm{B}_{1\mathrm{g}}}\cos(2\theta)+\Delta R{}_{\mathrm{B}_{2\mathrm{g}}}\sin(2\theta).\label{eq:poldepsimpl}
\end{equation}

Thus, in principle, by measuring the angle dependence of $\Delta R(\theta)$
and using Eq. (\ref{eq:poldepsimpl}) we can separate the $T$-dependent
dynamics of components associated with different symmetries and consequently
identify the states involved
\footnote{Note that this symmetry analysis is valid also within the photoinduced
absorption (PIA) picture commonly considered until now.}.

Previous analysis in the cuprates have indicated that the electronic
Raman scattering in the B$_{{\rm 2g}}$ symmetry probes excitations
in the nodal ($\pi$/2,$\pi$/2) direction in the $k$-space, while
the B$_{{\rm 1g}}$ scattering probes excitations in the antinodal
directions ($\pi$/2,0) and (0,$\pi$/2) \cite{Sugai00,Nemetschek97,Devereaux07}.
An $A_{1g}$ symmetry component is also present, whose origin is still
highly controversial \cite{Devereaux07}. Recent studies suggest the
presence of the PG in the nodal direction \cite{Sakai13} implying
an $s$-wave symmetry, in contrast to the common assumption of a PG
with nodes, indicating that the PG symmetry is still an open issue.
A detailed symmetry analysis of optical Pp experiments can therefore
potentially give important new information on the symmetry, lifetime
and temperature-dependence of nodal and anti-nodal excitations in
the cuprates and other superconductors with an enhanced bulk sensitivity
with respect to the time-resolved ARPES \cite{Perfetti07,Graf11}.

The optical measurements were performed on freshly cleaved slightly
overdoped (OD, $T_{\mathrm{c}}\approx$82 K) and underdoped (UD, $T_{\mathrm{c}}\approx$69
K) Bi2212 single crystals grown by the traveling solvent floating
zone method. 
For optimal signal-to noise ratio we used a pump at $E_{{\rm pu}}=$3.1
eV ($\lambda_{{\rm pu}}=$400 nm) and probe at $E_{{\rm pr}}=$1.55
eV ($\lambda_{{\rm pr}}=$800 nm) from a cavity-dumped Ti:sapphire
oscillator with a 120 fs pulses and a repetition rate of 270 kHz (to
avoid heating). The pump and probe beams were coaxially overlapped
by a dichroic mirror and focused to $20\mu$m diameter spot on the
$ab$-plane of the crystal with an objective lens ($f=$40 mm). Low-pass
filters were used to suppress any remaining scattered pump beam. We
use the notation where $x$ and $y$ point along the Cu-O bonds (Fig.
1(b)). The sample orientation was checked by x-ray diffraction, in
which the $b$-axis is determined by the direction of the multiple
peaks responsible for a one-dimensional (1D) superlattice modulation. 

To highlight the main findings of the study, we will concentrate on
the presentation of the OD sample, while noting that the results on
the UD sample are qualitatively similar. First we note that $\Delta R$
is found to be independent of the pump polarization to less than $\sim2\%$,
while the probe polarization dependence of $\Delta R$ is very temperature
dependent. 
\begin{figure*}[htb]
\includegraphics[width=0.9\textwidth]{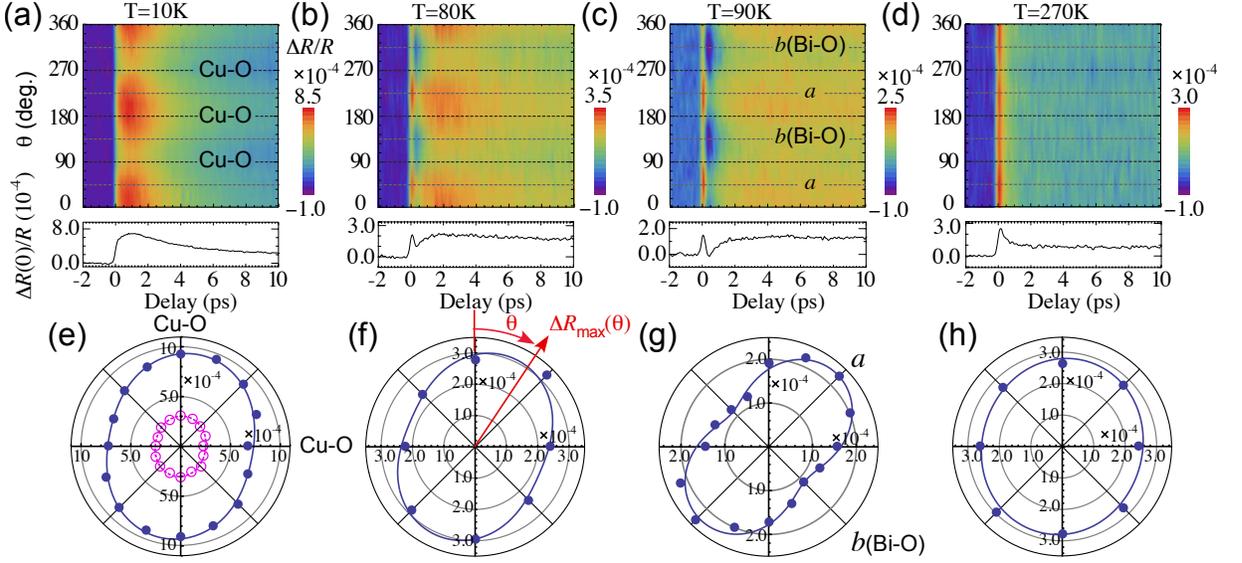} \caption{(a)--(d) $\Delta R(\theta)/R$ transients at typical temperatures.
(e)-(h) Polar plots of the maximum values of $\Delta R(\theta)/R$.
The solid lines indicate fits using Eq. (\ref{eq:poldepsimpl}). $\Delta R(\theta)$/$\Delta R$
at delay time of 10 ps is also shown (open circles with dashed fitting
line) in (e). All data were obtained from the OD sample. Note that
the Cu-O bonds directions are drawn horizontal and vertical, while
the crystalline axes are along the Bi-O bonds, and are rotated nearly
45$^{\circ}$ from the Cu-O bonds. }

\label{fig_pol} 
\end{figure*}

The angular dependencies of $\Delta R(t)$ at selected temperatures,
obtained by rotating the probe polarization from $\theta=0$ to $360^{\circ}$
at each temperature, are presented in Fig.\ref{fig_pol}. The upper
panels (a)-(d) show the intensity plots of $\Delta R(\theta)$ together
with the cross-sectional views at $\theta=0{}^{\circ}$. The polar
plots (e)-(h) show the angular dependence of the signal amplitude
$\Delta R(\theta)$. 

At the lowest temperature, where the SC signal is dominant, $\Delta R(\theta)$
is slightly elliptic, with the long axis close to, but not coincident,
with the Cu-O bonds direction (Fig.\ref{fig_pol} (a) and (e)). The
amplitude of the signal increases with increasing $\mathcal{F}$,
showing saturation behavior near $\mathcal{F}_{th}^{SC}=$1 $\mu$J/cm$^{2}$ \footnote{Note that the threshold is different than in Ref. \cite{Toda11} due
to a different pump wavelength.}.
With increasing $\mathcal{F}$, an additional fast relaxation signal
with opposite sign with respect to the SC signal appears, which persists
above $T_{\mathrm{c}}$ and disappears around $T^{*}$. This component
has been previously assigned to the PG QPs \cite{Liu08,Toda11}, where
$T^{*}\simeq$ 140 K for OD and 240 K for UD samples, respectively,
and is consistent with previous measurements \cite{Nakano98}. The
reason for the PG signal appearing at higher $\mathcal{F}$ is that
the PG component has a higher saturation threshold than the SC signal,
and therefore becomes visible below $T_{\mathrm{c}}$ with increasing
$\mathcal{F}$ \cite{Toda11}. 

In the PG state above $T_{c},$ but below $T^{*}$, the long axis
is oriented along the crystalline axes ($\theta\simeq45^{\circ}$)
(Fig.\ref{fig_pol} (c) and (g)). 

Above $T^{*}$, a signal with opposite sign to the PG (the same sign
as SC) becomes visible, which has been attributed to the electron
energy relaxation in the metallic state\cite{Gadermaier}. This high-temperature
signal is almost independent of $\theta$ (Fig.\ref{fig_pol} (d)
and (h)).

\begin{figure}[htb]
\includegraphics[width=0.98\columnwidth]{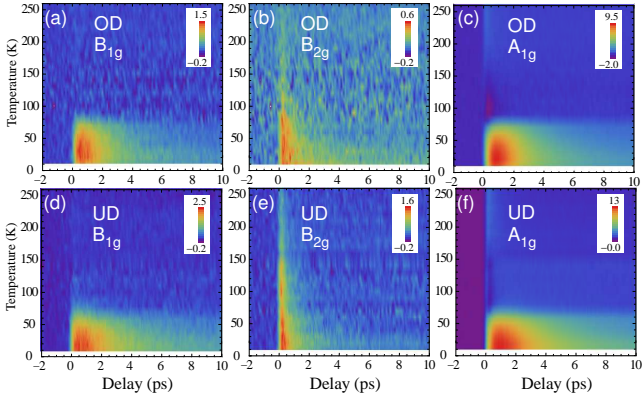} \caption{$T$-dependences of the B$_{{\rm 1g}}$, B$_{{\rm 2g}}$ and A$_{{\rm 1g}}$ components
of $\Delta R$ corresponding to SC, PG and metallic state relaxation
respectively for OD (top) and UD (bottom) samples; 
 The values of the color bars indicate $\Delta R/R\times10^{4}$.}

\label{fig_poltemp} 
\end{figure}

\begin{figure}[htb]
\includegraphics[bb=30bp 0bp 800bp 600bp,width=1\columnwidth]{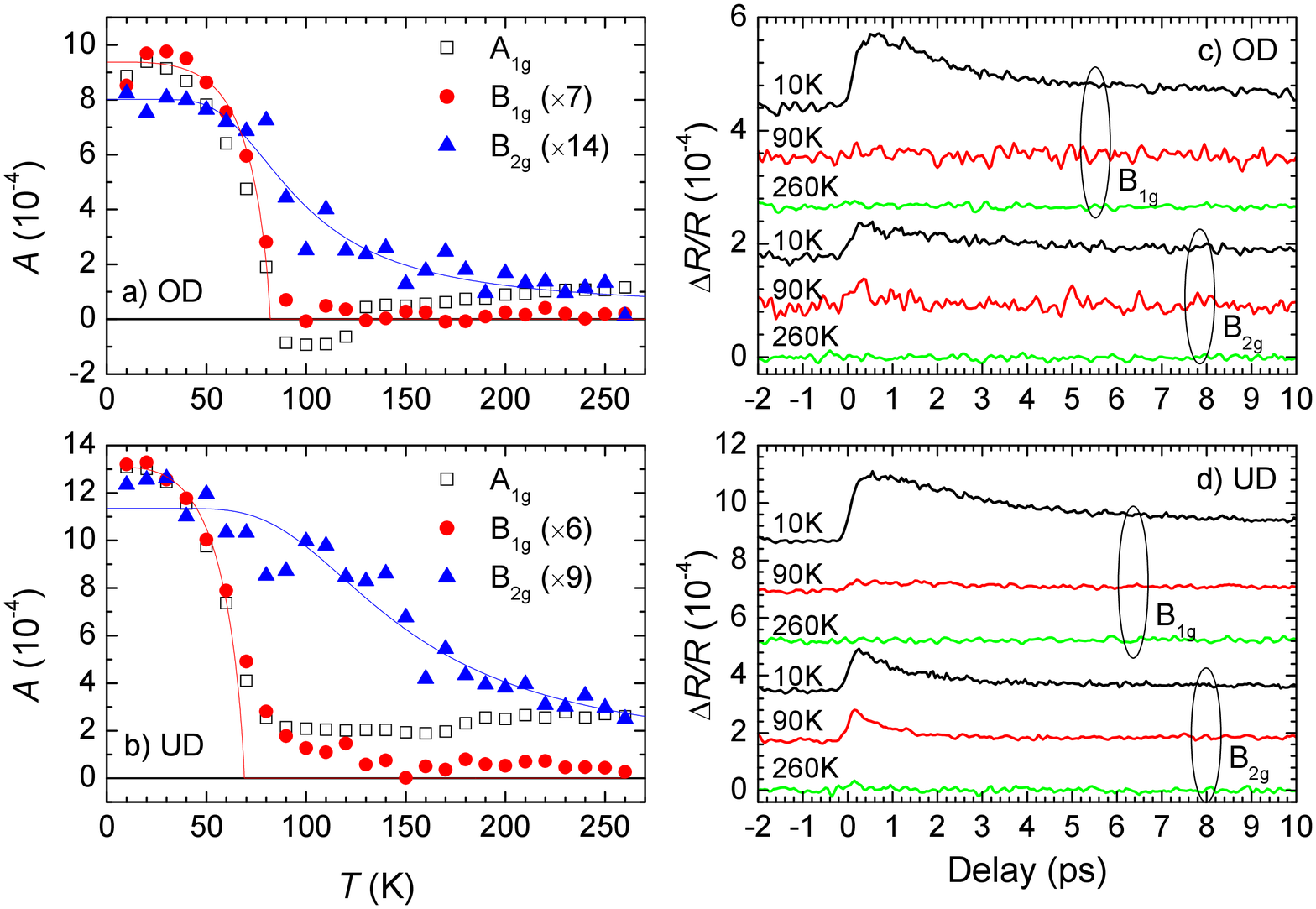}
\caption{$T$-dependences of the A$_{{\rm 1g}}$, B$_{{\rm 1g}}$ and B$_{{\rm 2g}}$ component
amplitudes for UD and OD samples. Note that all three components begin
to show an increase of the amplitude below $T^{*}$ in the UD sample.
In the case of the $A_{1g}$ symmetry the PG contribution is superimposed
on top of the nearly temperature-independent signal, and has a negative
sign. The solid and dashed lines are fits using Mattis-Bardeen \cite{MerteljKusar2010}
and Kabanov \cite{Kabanov99} models, respectively.}

\label{fig_scpg} 
\end{figure}

In Fig. \ref{fig_poltemp}, we present the $T$-dependences of the
$\Delta R^{\mathrm{B_{1g}}}/R$, $\Delta R^{\mathrm{B_{2g}}}/R$ and
$\Delta R^{\mathrm{A_{1g}}}/R$ obtained by fitting Eq. (\ref{eq:poldepsimpl})
to experimental data. $\Delta R^{\mathrm{B_{1g}}}$, and $\Delta R^{\mathrm{B_{2g}}}$
show clear dominance of the SC and PG responses, respectively. $\Delta R^{\mathrm{B_{1g}}}$
is only visible below $T_{\mathrm{c}}$ vanishing sharply at $T_{\mathrm{c}}$.
On the other hand, $\Delta R^{\mathrm{B_{2g}}}$ shows a gradual decrease
with increasing the temperature across $T_{\mathrm{c}}$ and a faster
sub-picosecond relaxation time, which is consistent with the general
behavior observed for the PG QPs \cite{Demsar99,Kabanov99}. The difference
of the $T$-dependences between OD and UD samples reflects the systematic
variation of the gaps with the doping level. 

In Fig. \ref{fig_scpg}, we plot the amplitudes of different components
for both OD and UD samples as a function of temperature. $\Delta R^{\mathrm{B_{1g}}}$
and $\Delta R^{\mathrm{A_{1g}}}$ show dominant intensity below $T_{c}$,
and their $T$-dependences can be fit well using the Mattis-Bardeen
formula \cite{MattisBardeen,Mertelj10}. The $\Delta R^{\mathrm{B_{2g}}}$
component can be fit well by the Kabanov's relaxation model \cite{Kabanov99,Kabanov05}
which gives a $T$-independent $\Delta_{{\rm PG}}=$30 meV for OD
and $\Delta_{{\rm PG}}$=41 meV for UD samples. The values of $\Delta_{{\rm PG}}$
in each sample are consistent with the values obtained from other
experiments \cite{Oda97,Kurosawa10}.

While the B$_{{\rm 1g}}$ component includes a signal extending to $\sim T^{*}$
in the UD sample, the B$_{{\rm 2g}}$ component does not show any measurable
change at $T_{c}$ within the noise level.

The absence of any pump polarization anisotropy has important consequences.
It rules out SRE as the excitation mechanism for any non-totally-symmetric
modes. At the same time, the fact that the B$_{{\rm 1g}}$ and B$_{{\rm 2g}}$
responses are observed by the probe means that they are somehow excited
by the pump pulse. Since DE can excite coherently only symmetric modes
we are left with the only possibility that the coherent B-symmetry
modes are excited because the underlying tetragonal point group symmetry
is spontaneously broken below $T^{*}$. 

Formally, the B$_{{\rm 2g}}$ symmetry breaking of the pseudotetragonal
symmetry is already present at room temperature in Bi2212 due to the
weak inherent orthorhombicity of the underlying crystal structure
from the BiO chain modulation arising from the mismatch of Bi-O and
Cu-O layers \cite{Miles1998}. In the resulting D$_{2h}$ point group
symmetry the $a$ and $b$ axes are rotated at $\sim45\text{\textdegree}$
with respect to the Cu-O bonds. The presence of any coherent B$_{{\rm 1g}}$
symmetry excitation, on the other hand, requires breaking of both
the CuO$_{2}$-plane pseudotetragonal ($\mathrm{D}_{4h}$) and the
Bi2212 $\mathrm{D}_{2h}$ symmetry
\footnote{In the $45\text{\textdegree}$ rotated Bi-2212 D$_{2h}$ point symmetry
the pseudotetragonal D$_{4h}$ B$_{1g}$ representation corresponds
to the B$_{1g}$ D$_{2h}$ representation. A B$_{1g}$ symmetry breaking
leads at least to lowering of the point symmetry from D$_{2h}$ to
C$_{2\mathrm{v}}$.
}.

In our data however, both symmetry breakings are suppressed at the
room temperature appearing clearly below $T^{*}$, implying that the
$\mathrm{B}{}_{2g}$ component is not simply a consequence of the
underlying Bi2212 orthorhombicity. The data in Fig. 4 clearly show
symmetry-breaking to occur near $T^{*}$ for both asymmetric components:
B$_{{\rm 1g}}$ and B$_{{\rm 2g}}$ in the UD sample while in the OD sample the
B$_{{\rm 1g}}$ component is clearly visible only in the superconducting
state. Furthermore, while the BiO chain ordering discussed above can
in principle cause the B$_{{\rm 2g}}$ symmetry breaking effect
it \emph{cannot} cause the observed B$_{{\rm 1g}}$ symmetry breaking
neither above nor below $T_{c}$.

Despite the $d$-wave SC order parameter corresponds to the B$_{{\rm 1g}}$
symmetry the observed effect can not be linked directly to the symmetry
of the order parameter. The SC order parameter is complex so any expansion
of the dielectric constant in terms of the SC order parameter can
only contain powers of $|\Delta_{\mathrm{SC}}|^{2}$, that are of
the A$_{1g}$ symmetry. This explains also the strong response of
the SC state in the A$_{1g}$ channel. 

The $\mathrm{B_{1g}}$ symmetry breaking can therefore only be associated
with an underlying order oriented along the Cu-O bonds. This finding
is consistent with previous STM measurements indicating the presence
of stripe order oriented along one of the two orthogonal Cu-O bond
directions on the crystal surface \cite{Vershinin04,Howald,Kohsaka07,Lang02}.
The magnitude of the B$_{1g}$ component significantly exceeds 10\%
of the A$_{1g}$ magnitude. Since the optical probe penetration depth
is of order of 100 nm \cite{Toda11} this indicates that the stripe
order and the B$_{1g}$ component are not limited to the surface and
are present also in the bulk. The sensitivity of the B$_{1g}$ component
to the SC order also suggests that the instability towards formation
of the stripe order and the superconductivity are intimately connected.

The distinct absence of the SC response in the B$_{2g}$ channel
is consistent with the sensitivity of the corresponding Raman vertex
to the nodal ($\pi$/2,$\pi$/2) direction in the $k$-space, where
the SC gap has nodes. On the other hand, the presence of the PG response
in B$_{2g}$ channel indicates that the PG response can, at least
in part, be associated with the nodal quasiparticles.
Remarkably, this implies the absence of nodes in the PG, consistently
with recent Raman results \cite{Sakai13}.

Upon reduction of symmetry from tetragonal to orthorhombic each of
the B symmetry breakings of the 4-fold axis can occur in two equivalent
directions (e.g. along $x$ or along $y$). Since our experiment is
stroboscopic and averages over many pulses B$_{1g}$ and B$_{2g}$
channels do not average out only if there is an underlying symmetry
breaking persisting between subsequent pulses separated by 4 $\mu$s.
This can be imposed by extrinsic defect structure or strain. In the
case of Bi2212 it appears that that the underlying B$_{2g}$ channel
symmetry breaking can originate in the weak orthorhombicity of the
crystal, while the underlying B$_{1g}$ channel symmetry breaking
is of extrinsic origin amplified by the softness of the CuO$_{2}$
planes towards stripe ordering or similar textures.

We reiterate that the observed  $T$-dependent in-plane probe polarization anisotropy
of the photoinduced reflectivity in Bi2212, and the concurrent \emph{absence}
of \emph{pump} polarization anisotropy rules out stimulated Raman
excitation as the external symmetry-breaking source. 

From this we conclude that the observed polarisation anisotropy indicates the presence of a bulk intrinsic spontaneous breaking of  both pseudotetragonal
CuO$_{2}$-plane (D$_{4h}$) and crystallographic Bi2212 (D$_{2h}$)
symmetry below $T^{*}$ in both underdoped and slightly overdoped
samples.

The detailed symmetry analysis of the Pp process in analysis of electronic
excitations presented here opens up new possibilities for investigating
the $k$-space anisotropy complementary to Raman spectroscopy and
new techniques like time-resolved ARPES.

\end{document}